\documentclass[12pt]{article}
\usepackage{scicite}
\usepackage{bm}
\usepackage{amsmath}
\usepackage{graphicx}

\newenvironment{sciabstract}{%
\begin{quote} \bf}
{\end{quote}}

\title{Chromodynamics of photons in an artificial non-Abelian magnetic Yang-Mills field}

\author {A. Fieramosca$^{1,2\dag}$, L. Polimeno$^{1,2\dag}$, G. Lerario $^{1\ast}$,
L. De Marco$^{1,2\ast}$,\\ 
M. De Giorgi$^{1}$, D. Ballarini$^{1}$, L. Dominici$^{1}$,
V. Ardizzone$^{1}$,\\
M. Pugliese$^{1}$, V. Maiorano$^{1}$, G. Gigli$^{1,2}$, C. Leblanc$^4$,  \\
G. Malpuech$^{4\ast}$, D. Solnyshkov$^{4,5\ast}$, D. Sanvitto$^{1}$\\
\\
\normalsize{$^{1}$CNR Nanotec, Institute of Nanotechnology,}\\
\normalsize{via Monteroni, 73100, Lecce, Italy.}\\
\normalsize{$^{2}$Dipartimento di Matematica e Fisica, Universit\'a del Salento,}\\
\normalsize{via Arnesano, 73100 Lecce, Italy.}
\\
\normalsize{$^{3}$Dipartimento di Fisica, Universit\'a degli Studi di Pavia, via
Bassi 6,}\\
\normalsize{27100 Pavia, Italy.}
\\
\normalsize{$^{4}$Institut Pascal, PHOTON-N2, Universit\'e Clermont Auvergne, CNRS,}\\
\normalsize{SIGMA Clermont, F-63000 Clermont-Ferrand, France.} \\
\normalsize{$^{5}$Institut Universitaire de France (IUF), 75231 Paris, France}\\
\\
\normalsize{$^\ast$To whom correspondence should be addressed}\\
\normalsize{E-mail: giovannilerario86@gmail.com, luisa.demarco@nanotec.cnr.it,}\\ \normalsize{dmitry.solnyshkov@uca.fr,malpuech@univ-bpclermont.fr}\\
\normalsize{$^\dag$ These authors contributed equally to this work.}
}

\date{}

\begin{document}

\baselineskip24pt

\maketitle

\begin{sciabstract}
Artificial gauge fields, simulating real phenomenologies that unfold in a vast variety of systems, offer extraordinary possibilities to study extreme physical effects in many different environments, from high energy physics to quantum mechanics and cosmology. They are also at the heart of topological physics. Here, exploiting a strongly anisotropic material under strong coupling regime, we experimentally synthesize a Yang-Mills non-Abelian gauge field acting on an exciton-polariton quantum flow like a magnetic field. We observe experimentally the corresponding curved trajectories and spin precession. This motion follows chromodynamics equations which normally describe the quarks strong interactions and their color. Our work therefore opens exciting perspectives of simulating quark-gluon dynamics using highly flexible photonic simulators. It makes of sub-atomic physics a potential new playground to apply topological physics concepts.
 \end{sciabstract}

\textbf{One Sentence Summary:} Cavity photons simulate a magnetostatic experiment with a non-Abelian field. 


Elementary bosons, mediators of the three fundamental forces of Nature, appear from the quantization of gauge fields. Historically, the first gauge theory was developed for the electromagnetic field and its quanta (photons). It is an Abelian gauge theory. The charges and the components of the vector potential are scalar and commute with each other. The next step was realized by Yang and Mills \cite{Yang1954} who introduced a non-Abelian gauge theory by replacing a scalar charge by a vector (the isospin). The components of the vector potential are the Pauli matrices forming the SU(2) group, which obviously do not commute \cite{Ryder}. Complemented by the Higgs mechanism \cite{Higgs1964}, non-Abelian Yang-Mills gauge theories describe both the strong and weak interactions, allowing to build the whole standard model of elementary particles \cite{Quigg1997}.

Since then, a new playground has been found for gauge theories with the invention and experimental implementation of emergent gauge fields \cite{aidelsburger2018artificial}. 
The most well-known example is the Berry curvature, which is often interpreted as a magnetic field analog, but defined in a parameter space \cite{berry1984quantal} (e.g. momentum space). In all cases, the evolution along a trajectory (either in parameter space or in real space) is associated with an additional geometric phase (Berry or  Aharonov-Bohm phase). The Berry phase can be seen as an emergent vector potential with scalar components, from which gauge-invariant quantities are constructed. Topological physics as a whole can therefore be viewed as emergent electrodynamics described by an Abelian gauge theory \cite{Hasan2010, lu2014topological,ozawa2019topological,Cooper2019}. Emergent \emph{non-Abelian} gauge theories, describing  analogs of the strong or electro-weak interactions, have been much less explored so far. 

A very promising strategy relies on the recently discovered mapping \cite{Jin2006,Tokatly2008} between the Rashba spin-orbit coupling (SOC) \cite{Rashba1984} for massive particles, well-known in solid state physics, and the Yang-Mills gauge field theory. This mapping opened the exciting perspective of developing analog chromodynamics in solid state systems. Indeed, the only real particles subject to non-Abelian Yang-Mills fields are quarks whose behavior is described by quantum chromodynamics. Such analogue simulations are especially important because the trajectories of the real quarks are impossible to track experimentally. It is important to distinguish different recent realizations of artificial Rashba-like SOCs. As a dominant term, it gives rise to a 2D Dirac Hamiltonian with non-commuting $x$ and $y$ projections, but without a non-Abelian gauge field. Such configurations have been implemented recently in cold atom systems \cite{Spielman2011,Wu2016}.
Only when the Rashba SOC applies to a massive particle in addition to the main kinetic energy term, it can be considered as a minimally-coupled  non-Abelian Yang-Mills gauge field.
A scheme to synthesize such Rashba-Dresselhaus SOC has been theoretically proposed for confined massive photons \cite{terccas2014non,Chen2019} and  only very recently implemented experimentally using planar microcavities in the strong exciton-photon coupling regime \cite{gianfrate2019direct} and filled with liquid crystals \cite{Rechcinska2019}.
The observation of a Yang-Mills monopole in a synthetic 4D space has also been reported \cite{Sugawa2018}.
 At the same time, other researchers attacked the problem of non-Abelian gauge fields from another side, by considering a non-Abelian version of the Aharonov-Bohm effect \cite{Wu1975}, \cite{Yang2019}. It consists in making successively precess the light polarisation pseudospin about two constant, but differently oriented effective magnetic fields. Indeed, the final pseudospin orientation depends on the ordering of the two constant fields. This ordering sensitivity is a consequence of the non-commutativity of Pauli matrices. Using the Yang-Mills formalism, these  terms are the \emph{time-like} components of the four-vector potential, affecting the phase and the interference of the particles, but not their real space trajectories.

In this work, we implement experimentally another type of Yang-Mills vector potential with \emph{space-like} components affecting the real space trajectories via an analog of the Lorentz force. This vector potential is deduced from the mapping to a Rashba-Dresselhaus Hamiltonian which emerges in a strongly-coupled optical system. The non-Abelian character of the Gauge field implies the existence of a non-zero  "magnetic" Yang-Mills field. We provide a direct measurement of the transverse acceleration of an exciton-polariton wave packet, caused by the resulting "magneto-static" Yang-Mills force, together with the corresponding spin precession, which for quarks corresponds to a color change. This coupled spatial and spin dynamics is successfully described by 2-color chromodynamics equations. This experiment therefore represents an SU(2) analog of quarks with a direct experimental accessibility.

\begin{figure}[tbp]
\centering
\includegraphics[width=0.8\linewidth]{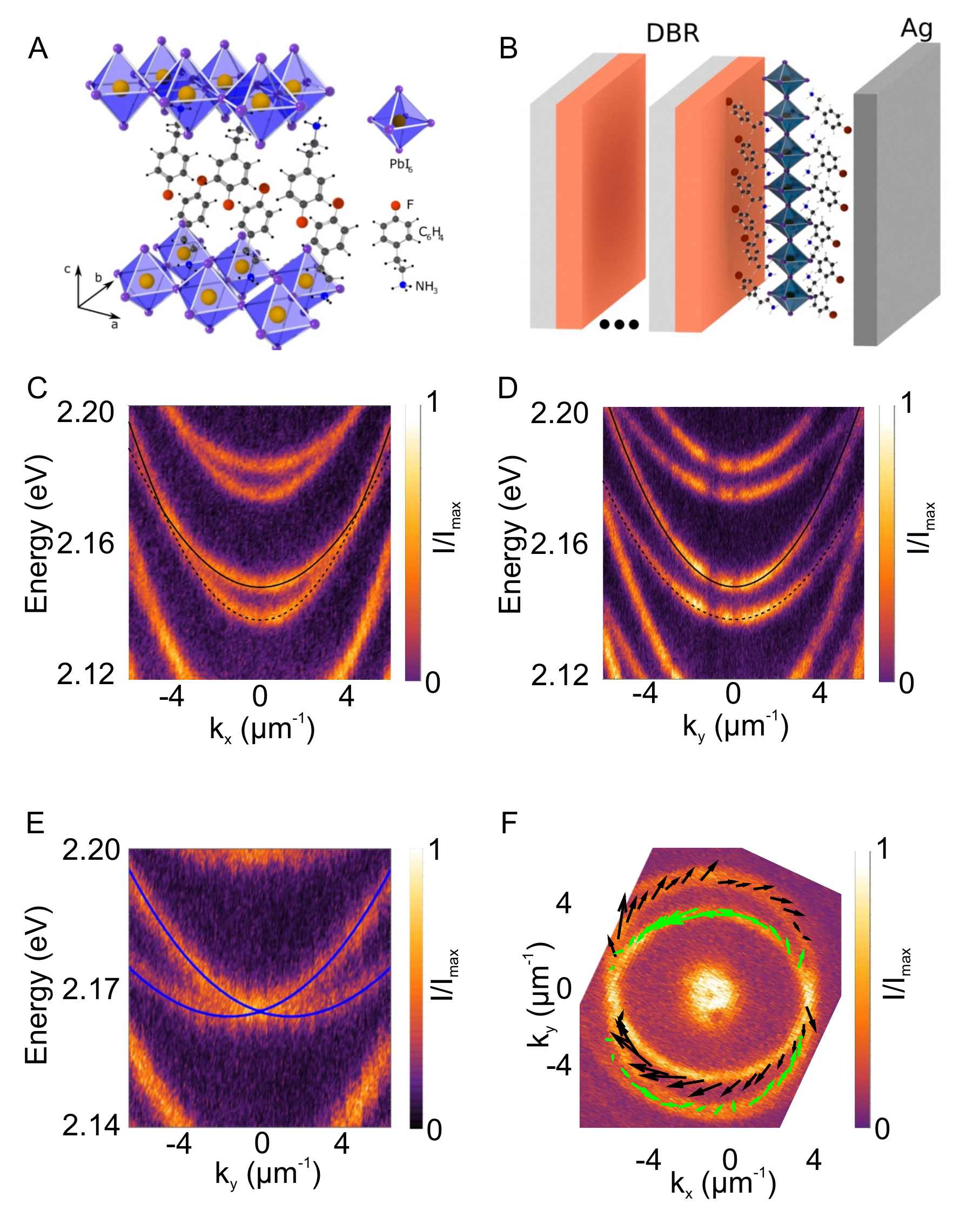}
\caption{\textbf{Experimental implementation of a non-Abelian gauge field.} A) Sketch of a $2D$ perovskite single crystal structure. B) Schematic representation of the microcavity sample. Perovskite flakes are embedded in an optical microcavity made by a DBR (seven $TiO_2/SiO_2$ pairs) and a $80~nm$ thick silver mirror. 
C,D) Experimental dispersions along $k_x$ and $k_y$ showing two diabolical points along $k_x$. E) Experimental dispersion along $k_y$ for a fixed value of $k_{x}=4.48 ~ \mu m^{-1}$ (crossing point) highlights the formation of a Rashba-type dispersion. Solid and dashed lines -- dispersion fit with the parameters given in the main text. F) PL at the diabolical point energy together with the pseudo-spin orientation (arrows) obtained from polarisation measurments. The monopolar pseudospin texture around the diabolical points is another signature of the Rashba SOC.
}
\end{figure}

The sample is based on a single crystal of 2D hybrid organic-inorganic
perovskite (Fig.~1A, see Methods) placed in an asymmetric planar
microcavity with a distributed Bragg reflector as a bottom mirror
and a $80~nm$-thick silver layer as top mirror (Fig.~1B). The active
material is 4-fluoro-phenethylammonium tetraiodoplumbate (henceforth
referred to as PEAI-F) a multiple quantum well system consisting of
$PbI_{2}$ layers sandwiched between organic insulator layers. The
perovskite layer thickness is $5~\mu m$. The quality factor of the
planar microcavities modes is $Q_{f}\simeq1000$. The wavevector (k)
of the bare light modes perpendicular to the mirrors is quantized,
giving rise to a series of 2D bands, with approximately parabolic
dispersion at $k\sim0$. The role of the \emph{charge vector} in our
case is played by the pseudospin associated with the polarization
of light (the Stokes vector, see methods). The spin-orbit coupling arises from
the energy splitting between the Transverse-Electric (TE) and Transverse-Magnetic
(TM) modes of the microcavity \cite{terccas2014non}. The polarized modes form two parabola
with different effective masses $m_{TM}$, $m_{TE}$. The perovskite
excitons located at 2.39~eV are strongly coupled
with the photonic modes of the cavity, forming exciton-polariton modes
(polaritons) at room temperature \cite{fieramosca2018tunable} with
a Rabi Splitting of 208~meV. Thanks to the presence of the fluorine,
the crystal symmetry of the perovskite provokes a strong linear birefringence. The birefringence breaks the cylindrical symmetry of the TE and
TM modes and lifts their degeneracy at $k=0$. Figures~1C and 1D
show the transmission of the cavity versus energy, for wave vectors
along the two perpendicular orientations $k_{x}$, $k_{y}$. Our microcavity
is relatively thick and the quantized polarisation doublets are close
to each other, separated by about 50~meV. One can observe a clear
increase of the effective masses for one polarisation doublet to another
while going to higher energy which is due to the increased excitonic
content of the corresponding exciton-polariton eigen modes. Within
a polarisation doublet, the behaviour is radically different along
the two wave vector orientations and can be understood by writing
down an effective Hamiltonian in the parabolic approximation on the
circular polarisation basis \cite{terccas2014non}:

\begin{eqnarray}
H_{\mathbf{k}}=\begin{pmatrix}
E_0+\frac{\hbar^2k^2}{2m}&\beta_0-\beta k^2 e^{-2i\varphi} \\
\beta_0-\beta k^2 e^{2i\varphi}& E_0+ \frac{\hbar^2k^2}{2m}
\label{Hameff}
\end{pmatrix} 
\end{eqnarray}
where $m=m_{TM}m_{TE}/(m_{TM}+m_{TE})$. $k=|\mathbf{k}|=\sqrt{k_x^2+k_y^2}$ is the in-plane wavevector ($k_x=k\cos{\varphi}$, $k_y=k\sin{\varphi}$, $\varphi$ is the propagation angle).$\beta_0$ is the optical birefringence. Along $k_x$, the X polarized mode corresponds to the TM mode with the smaller mass. The two parabola cross at $k^0_y=(\beta_0/\beta)^{1/2}=±4.48~\mu m^{-1}$ as shown in Fig~1C giving rise to a diabolical point. Along $k_y$, ($\varphi=\pi$), the off-diagonal terms add up. It means that the lowest polarisation mode at $k=0$ (polarised along X) corresponds to the TE mode with the larger mass: the splitting increases with $k_y$ (Fig.~1D). The fit of the experimental data with the effective Hamiltonian~\eqref{Hameff} yields  $E_0=2.1415~eV$, $m=2.4 \cdot 10^{-5}~m_e$, $ \beta = 2.5 \cdot 10^{-4} ~eV \mu m^{2}$, and $\beta_0= 10~meV$. 

Interestingly, around this diabolical point, the effective Hamiltonian can be rewritten \cite{terccas2014non,gianfrate2019direct} as a Rashba-like Hamiltonian \cite{Rashba1984}
\begin{equation}
    \hat{H}_R=\frac{1}{2m}\hat{\bm{p}}^2+\alpha \bm{\sigma}\cdot\hat{\bm{p}}=\frac{1}{2m}\left(\hat{\bm{p}}+m\alpha \bm{\sigma}\right)^2-m\alpha^2\sigma^0
    \label{hamR}
\end{equation}
where $\mathbf{\sigma}$ is a vector of Pauli matrices. ${p}=\hbar q$ is the momentum, $q=k-k^0_y$. $\alpha=\sqrt{\beta_0\beta}$. The Rashba-like nature of this diabolical point is experimentally proven in Figs.~1E and F. Fig.~1E  shows the dispersion along $k_y$ at $k_x=k^0_y$ which demonstrates the conical shape of the dispersion around the crossing point. Figure~1F shows the pseudospin distribution of the eigenstates at the energy of the crossing point. The pseudospin is deduced from the measured polarisation degree of emission (see Methods). One can see that the pseudospin is pointing in opposite directions at each side of the crossing, in agreement with the Rashba Hamiltonian picture. With respect to \cite{gianfrate2019direct}, the visibility of the crossing point, given by $k^0_y=(\beta_0/\beta)^{1/2}$   is here considerably enhanced because of the large birefringence imposed by the anisotropic perovskite crystal. 


Now let us write below the general non-relativistic Hamiltonian of a massive matter field minimally coupled with a Yang-Mills field determined by a vector potential $A_\mu^a$:
\begin{equation}
    H_{YM}=\frac{1}{2m}\left(\hat{\bm{p}}-\eta\bm{A}^a\sigma^a\right)^2+\eta A_t^a\sigma^a
    \label{HYM}
\end{equation}
which can be derived from the general Yang-Mills Lagrangian \cite{Yang1954,Ryder} (see Supplemental for details). Here, the coupling constant is $\eta=\hbar/2$ (the quantum of spin). We use upper number indices $0-3$ for Pauli matrices. Comparing this expression with Eq.~\eqref{hamR}, we see that only two components of the emergent vector potential are non-zero: $A_x^1=-m\alpha/\eta$, $
A_y^2=-m\alpha/\eta$. The emergent vector potential is constant, but because the underlying symmetry group is SU(2) and not U(1), a constant vector potential results in a non-zero field strength tensor with components given by $F_{\mu\nu}^a=\partial_\mu A_\nu^a-\partial_\nu A_\mu^a-\eta\varepsilon^{abc}A_\nu^b A_\mu^c$, where $\mu,\nu$ span $(t,x,y,z)$. The nonzero components of the field strength tensor read: $F_{yx}^3=-F_{xy}^3=-m^2\alpha^2/\eta$. These components couple to the spatial degrees of freedom, as discussed below. For comparison, the time-reversal symmetry breaking optical elements used in Ref.~\cite{Yang2019} correspond to the Zeeman splitting terms $\eta A_t^a\sigma^a$ of Eq.~\eqref{HYM}, that is, to the coupling of the spin with a constant external field, and not to a spin-orbit coupling. They define the time-like components of the vector potential, affecting the \emph{interference} of the particles, but not their \emph{trajectories}.

Similar to the Lorentz force, which is given by the product between the electric current and the field, the general form of the force (acceleration) provided by the Yang-Mills field links a unified spin-current vector $J$ and the field strength tensor $F$. The equations of motion for the velocity $\bm{v}$ and spin $\bm{s}$ a classical relativistic particle coupled to the Yang-Mills field read
\begin{equation}
    m\,dv^\mu/d\tau=\bm{J}_\nu\cdot\bm{F}^{\mu\nu},\quad d\bm{s}/d\tau=-\eta\bm{A}_\mu\times\bm{J}^\mu
    \label{chrome}
\end{equation}
where $\bm{J}_\nu=\bm{s}v_\nu$ is the spin current. 
These two equations are solved together to find the particle trajectory and color dynamics in classical chromodynamics \cite{Boozer2011} studying generalized vector charges (in contrast with electrodynamics and its scalar charges). We note that chromodynamics equations describing quark and gluon dynamics are  based on SU(3) gauge theory and involves three colors (spin-1), whereas the present implementation is SU(2), as in the original Yang-Mills paper \cite{Yang1954}, and involves two colors only (spin-$1/2$).

In the specific case we consider here, the acceleration is given by $a_x=-4m\alpha^2J_y^3/\hbar^2$, $a_y=4m\alpha^2J_x^3/\hbar^2$, where $J_x^3$, $J_y^3$ are the circular (spin-up/down) components of the polariton spin current propagating along $x$ and $y$ respectively. The YM acceleration is therefore transverse, but acts on the circular component of the spin current, instead of affecting the charge current for the Lorentz force. The magnitude of the YM force is given by $\alpha^2=\beta_0\beta/2$. It is therefore expected to be 3 orders of magnitude larger than in GaAs-based microcavities, for instance \cite{gianfrate2019direct}, thank to the enhanced anisotropy of PEAIF-F (high $\beta_0$). A particularly important advantage of studying the non-Abelian Yang-Mills field with polaritons as compared to electron-based spintronics is that photons are neutral and do not couple to the electromagnetic field, contrary to the electrons, which allows the YM-related effects to be the dominant ones. We stress that the effect of the "magnetic" Yang-Mills field we consider here differs crucially from a simple spin-dependent Lorentz force \cite{Beeler2013}. In the latter case, the equations for the two spin components are decoupled, and the second equation of chromodynamics~\eqref{chrome}  does not play any role: colors do not change and the Lorentz force is constant.

\begin{figure}[tbp]
\includegraphics[width=0.9\linewidth]{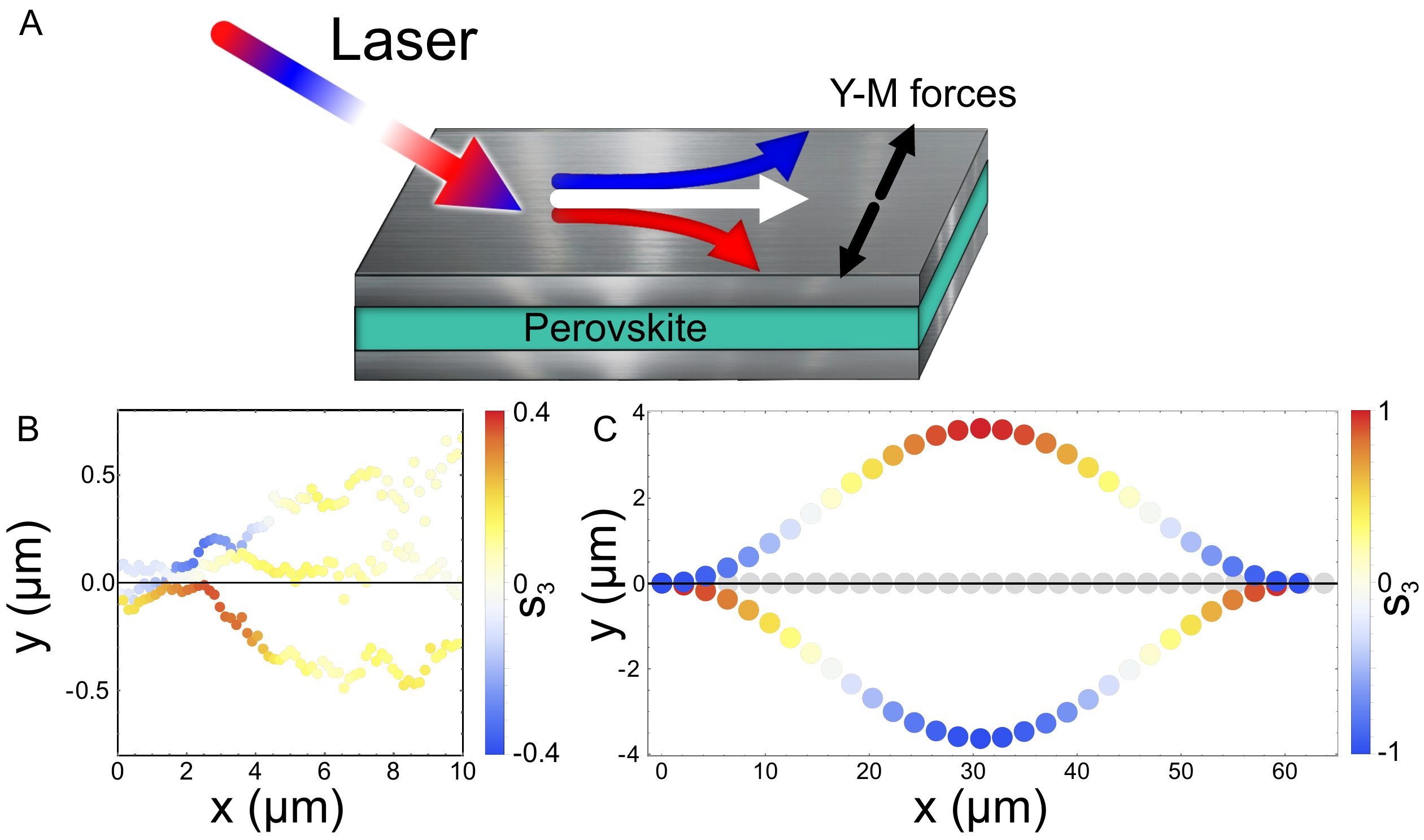}
\caption{Chromodynamics in polariton propagation. A) Scheme of the experiment. A polarized laser creates a flow which is deviated by the Yang-Mills field depending on the sign of the spin current. Polarization of the excitation: left-circular ($s^3=-1$), vertical ($s^1=1$, $s^3=0$), right-circular ($s^3=1$). B) Experimentally measured center of mass trajectories for three excitation conditions ($s^3=-1,0,1$). Dot color corresponds to $s^3$. C) chromodynamic simulation of propagation trajectories and color for the same initial conditions ($s^3=-1,0,1$). Dot color corresponds to $s^3$.
}
\end{figure}

In order to demonstrate experimentally this transverse acceleration, we use a unique specificity of cavity polaritons: the possibility to resonantly create a wavepacket with a well defined pseudospin, centered on a specific  state in reciprocal space, and then to study its real space evolution. The first experiment we perform consists in exciting resonantly the vicinity of the Rashba diabolical point  with a pulsed polarized laser, which in the language of Yang-Mills chromodynamics corresponds to creating a color current. The scheme of the experiment is shown in Fig.~2A.  Figure~2B shows the center of mass of the intensity distribution versus the propagation direction extracted from the experiment together with the wave packet spin (color) for 3 excitation conditions: $\sigma^+$, $\sigma^-$, and vertical polarization ($s^3=1,0,-1$ with $s^1=0,1,0$). The two colors (red and blue) correspond to the 2 $s^3$ spin components ($\sigma^+$ and $\sigma^-$). Figure~2C shows the trajectories and colors calculated theoretically with the equations of chromodynamics~\eqref{chrome} using the parameters $m, \alpha, \beta, \beta_0$ extracted from the experiments and given above. The Yang-Mills magnetic field acts on the color currents, which exhibit lateral deviation depending on their color and velocity. At the same time, the color itself changes depending on the propagation direction. This gives rise to opposite oscillating trajectories for red ($\sigma^+$) and blue ($\sigma^-$) wave packets. Both effects are absent for colorless excitation (vertical polarization): the wave packet propagates along a straight line. This behaviour is clearly visible on the 10 $\mu m$ propagation distance accessible experimentally and which is limited by cavity Q-factor and sample inhomogeneities. These measurements are in excellent agreement with the chromodynamics simulations which are presented on a larger scale in Fig.~2C, showing a full oscillation period.  We note that other, qualitatively different types of behavior are possible for other parameters, including closed circular orbits, analogues of cyclotron orbits or Landau levels in an ordinary Abelian magnetic field. These represent an interesting subject for future studies.

The second experiment consists in creating an energy potential in the plane of the cavity and to launch the flow of neutral particles (vertically polarized polaritons, $s^1=1$, $s^3=0$) against the defect, as shown in Fig.~3A. This type of experiment has previously been used in microcavities in the high density regime. It allowed to demonstrate polariton superfluidity \cite{Amo2009,lerario2017room}, the formation of oblique solitons \cite{Amo2011}, half-solitons\cite{Hivet2012}, and of vortex anti-vortex pairs \cite{Dominici2018}. Here we work in the linear, low density regime, but in presence of a non-Abelian Yang-Mills gauge field. 

\begin{figure}[tbp]
\includegraphics[width=0.9\linewidth]{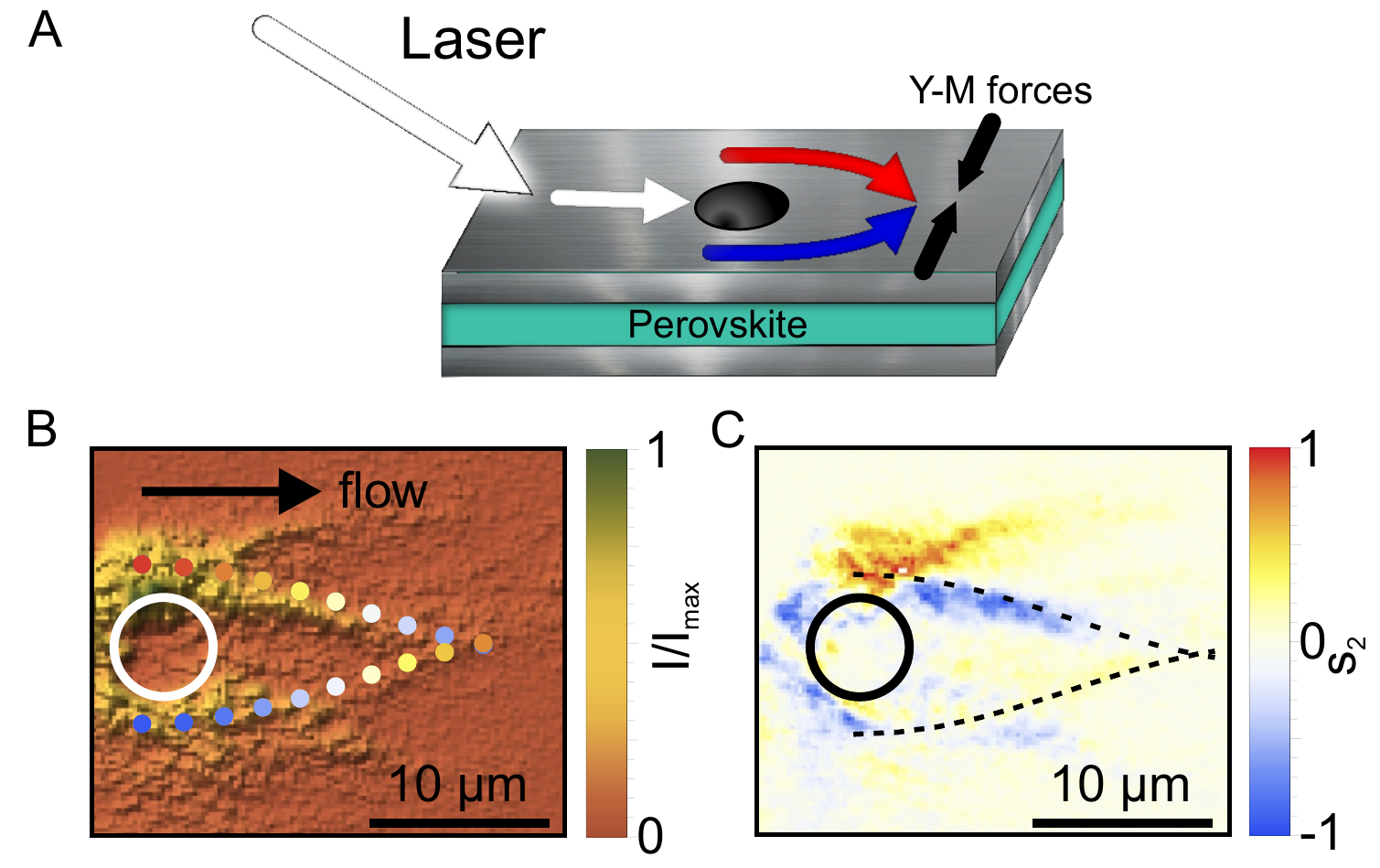}
\caption{Chromodynamics behind a defect potential.   A) Scheme of the experiment. A linear polarized laser creates a propagating flow, which hits a potential and splits into circular-polarized flows deviated by the Yang-Mills field. B) Experimental image of the total emission intensity (false color). Colored dots show the theoretical results (color corresponds to the $s^3$ spin projection, the scale is the same as on Fig. 3). The white circle shows the position of the potential defect. C) Experimental image of the difference between the diagonal polarization intensities ($s^2$). Dashed lines mark the theoretical propagation trajectories.\label{fig4}
}
\end{figure}

The polariton flow is in different conditions upstream and downstream of the defect. Upstream, the particle trajectory is strongly constrained by the defect potential and by the quantum pressure. The effects of the magnetic Yang-Mills force on trajectories is negligible compared to these two other contributions. However, the effects of the Yang-Mills field on the spin evolution, described by the second equation of chromodynamics \eqref{chrome}, are not negligible. For upward and downward propagation, the gained $s^3$ components, proportional to $J_y^1$, have opposite signs. Opposite colors are thus generated above and below the defect, as shown in the sketch Fig.~3A. In the region after the defect, the magnetic Yang-Mills force becomes dominant, and the coupled chromodynamics equations~\eqref{chrome} completely describe the particle trajectories and their spins. Due to the opposite colors (and therefore spin currents $J_x^3$) above and below the defect, the transverse force in Eq.~\eqref{chrome} is also opposite. This force brings the particles into the shadow of the defect where a convergent flow should form. This convergent flow is clearly visible in Fig.~3B showing the experimentally measured spatial images of the total particle density.The results of chromodynamic simulations based on Eqns.~\eqref{chrome} corresponding to the wave packet trajectories are shown as points, whose color shows the $s^3$ spin projection. Figure~3C presents the  difference between the $s^2$ components (chosen as the new color basis), showing the best contrast due to the particular color dynamics. Dashed lines show the calculated particle trajectories (same as in panel B). It confirms that the density flows observed in total intensity in Fig.~3B are not due to a particular disorder pattern, but to the chromodynamics in presence of a single well-defined defect. We would like to underline that these curved trajectories have nothing in common with the recently observed anomalous Hall drift \cite{gianfrate2019direct} of accelerated polariton wave packets which is induced by the non-zero Berry curvature of the polariton bands when time-reversal symmetry is broken. Anomalous Hall effect caused by an emergent Abelian magnetic field in the \emph{reciprocal} space occurs during the adiabatic motion of a wave packet within a \emph{single} band. On the opposite, the non-Abelian magnetic field in the present work acts in \emph{real} space, and the oscillating trajectories in Figs.~2 and 3  are due to beatings between \emph{two} coherently excited eigenstates (spin precession described by the second equation of chromodynamics).

In particle physics, the particles coupled with a non-Abelian Yang-Mills gauge field are the quarks, it is their color that determines their coupling to the field, and the excitations of the field are the gluons. In our analogue system, the field is constant and fixed externally, so there are no analogue gluons in this experiment yet. This allows to obtain a simpler, fully solvable configuration, with the enormous advantage of the possibility of direct experimental measurements of the particle trajectories and of their isospin orientation which is evidently impossible in the original quark-gluon system. In the future, the strong polariton-polariton interactions, which are moreover spin-anisotropic, can be used to extend the analogy to describe not only the "magnetostatic" limit of chromodynamics, but to gain access to the whole range of classical and quantum chromodynamic phenomena.


\textbf{Acknowledgements:}
The authors acknowledge Paolo Cazzato for technical support, Iolena Tarantini for the metal evaporation, Sarah Porteboeuf-Houssais and Jean Orloff for useful discussions about quantum chromodynamics. \textbf{Funding:} We acknowledge the ERC project ElecOpteR grant number $780757$ and the project \textquotedblleft TECNOMED - Tecnopolo di Nanotecnologia e Fotonica per la Medicina di Precisione\textquotedblright , (Ministry
of University and Scientific Research (MIUR) Decreto Direttoriale n. $3449$ del $4/12/2017$, CUP $B83B17000010001$). G.G. gratefully acknowledges the project PERSEO-PERrovskite-based Solar cells: towards high Efficiency
and lOng-term stability (Bando PRIN $2015$-Italian Ministry of University and Scientific Research (MIUR) Decreto Direttoriale $4$ novembre $2015 n. 2488$, project number $20155LECAJ$).
We also acknowledge the support of the project "Quantum Fluids of Light"  (ANR-16-CE30-0021), of the ANR Labex Ganex (ANR-11-LABX-0014), and of the ANR program "Investissements d'Avenir" through the IDEX-ISITE initiative 16-IDEX-0001 (CAP 20-25).
\textbf{Competing interests:} Authors declare no competing interests. \textbf{Data and materials availability:} The datasets generated and analyzed during the current study are available via the Open Science Framework (OSF) repository. The materials are available upon reasonable request.

\bibliographystyle{Science}
\bibliography{Bibliography}

\begin{thebibliography}{10}

\bibitem{Yang1954}
C.~N. Yang, R.~L. Mills, {\it Phys. Rev.\/} {\bf 96}, 191 (1954).

\bibitem{Ryder}
L.~H. Ryder, {\it Quantum Field Theory: Second Edition\/} (Cambridge University
  Press, Cambridge, 1996).

\bibitem{Higgs1964}
P.~W. Higgs, {\it Phys. Rev. Lett.\/} {\bf 13}, 508 (1964).

\bibitem{Quigg1997}
C.~Quigg, {\it Gauge theories of the strong, weak, and electromagnetic
  interactions\/} (Addison-Wesley, New York, 1997).

\bibitem{aidelsburger2018artificial}
M.~Aidelsburger, S.~Nascimbene, N.~Goldman, {\it Comptes Rendus Physique\/}
  {\bf 19}, 394 (2018).

\bibitem{berry1984quantal}
M.~V. Berry, {\it Proceedings of the Royal Society of London. A. Mathematical
  and Physical Sciences\/} {\bf 392}, 45 (1984).

\bibitem{Hasan2010}
M.~Z. Hasan, C.~L. Kane, {\it Rev. Mod. Phys.\/} {\bf 82}, 3045 (2010).

\bibitem{lu2014topological}
L.~Lu, J.~D. Joannopoulos, M.~Solja{\v{c}}i{\'c}, {\it Nature Photonics\/} {\bf
  8}, 821 (2014).

\bibitem{ozawa2019topological}
T.~Ozawa, {\it et~al.\/}, {\it Reviews of Modern Physics\/} {\bf 91}, 015006
  (2019).

\bibitem{Cooper2019}
N.~R. Cooper, J.~Dalibard, I.~B. Spielman, {\it Rev. Mod. Phys.\/} {\bf 91},
  015005 (2019).

\bibitem{Jin2006}
P.-Q. Jin, Y.-Q. Li, F.-C. Zhang, {\it J. Phys. A: Math. Gen.\/} {\bf 39}, 7115
  (2006).

\bibitem{Tokatly2008}
I.~V. Tokatly, {\it Phys. Rev. Lett.\/} {\bf 101}, 106601 (2008).

\bibitem{Rashba1984}
Y.~A. {Bychkov}, {\'E}.~I. {Rashba}, {\it Soviet Journal of Experimental and
  Theoretical Physics Letters\/} {\bf 39}, 78 (1984).

\bibitem{Spielman2011}
Y.-J. Lin, K.~Jimenez-Garcia, I.~B. Spielman, {\it Nature\/} {\bf 471}, 83
  (2011).

\bibitem{Wu2016}
Z.~Wu, {\it et~al.\/}, {\it Science\/} {\bf 354}, 83 (2016).

\bibitem{terccas2014non}
H.~Ter{\c{c}}as, H.~Flayac, D.~Solnyshkov, G.~Malpuech, {\it Physical Review
  Letters\/} {\bf 112}, 066402 (2014).

\bibitem{Chen2019}
Y.~Chen, {\it et~al.\/}, {\it Nature Comm.\/} {\bf 10}, 3125 (2019).

\bibitem{gianfrate2019direct}
A.~{Gianfrate}, {\it et~al.\/}, {\it arXiv e-prints\/} p. arXiv:1901.03219
  (2019).

\bibitem{Rechcinska2019}
K.~Rechci{\'n}ska, {\it et~al.\/}, {\it Science\/} {\bf 366}, 727 (2019).

\bibitem{Sugawa2018}
S.~Sugawa, F.~Salces-Carcoba, A.~R. Perry, Y.~Yue, I.~B. Spielman, {\it
  Science\/} {\bf 360}, 1429 (2018).

\bibitem{Wu1975}
T.~T. Wu, C.~N. Yang, {\it Phys. Rev. D\/} {\bf 12}, 3845 (1975).

\bibitem{Yang2019}
Y.~Yang, {\it et~al.\/}, {\it Science\/} {\bf 365}, 1021 (2019).

\bibitem{fieramosca2018tunable}
A.~Fieramosca, {\it et~al.\/}, {\it ACS Photonics\/} {\bf 5}, 4179 (2018).

\bibitem{Boozer2011}
A.~D. Boozer, {\it American Journal of Physics\/} {\bf 79}, 925 (2011).

\bibitem{Beeler2013}
M.~C. Beeler, {\it et~al.\/}, {\it Nature\/} {\bf 498}, 201 (2013).

\bibitem{Amo2009}
A.~Amo, {\it et~al.\/}, {\it Nature Phys.\/} {\bf 5}, 805 (2009).

\bibitem{lerario2017room}
G.~Lerario, {\it et~al.\/}, {\it Nature Physics\/} {\bf 13}, 837 (2017).

\bibitem{Amo2011}
A.~Amo, {\it et~al.\/}, {\it Science\/} {\bf 332}, 1167 (2011).

\bibitem{Hivet2012}
R.~{Hivet}, {\it et~al.\/}, {\it Nature Physics\/} {\bf 8}, 724 (2012).

\bibitem{Dominici2018}
L.~Dominici, {\it et~al.\/}, {\it Nature Comm.\/} {\bf 9}, 1467 (2018).

\end{thebibliography}

\section{Supplementary Materials}

 \subsection{Materials and Methods}

\underline{Synthesis of $2D$ perovskite flakes} 

PEAI-F solutions are prepared in a nitrogen-filled glovebox by dissolving 230 mg of lead iodide and 267 mg of $4$-fluorophenethylammonium iodide in $1$~mL of $\gamma$-butyrolactone and stirring at 70$^{\circ}$C for $30$ minutes. $2~\mu$l of the perovskite solution are deposited on the top of a DBR, immediately covered by the second DBR substrate and left undisturbed for 10-12 hours in a saturated environment of dichloromethane (antisolvent). At the end of the process, large area (millimeter size) and a few microns thick single crystals appear on top of the DBR.
\\

\underline{Microcavity Sample Fabrication}

DBRs are made by seven pairs of TiO$_2$/SiO$_2$ ($63$~nm/$94$~nm) deposited by radio-frequency (RF) sputtering process -- in
an Argon atmosphere under a total pressure of $6\cdot10^3$~mbar and at RF power of $250$~W -- on top of a $1$~mm glass substrate. The perovskite single crystals are grown on the DBR (see above) and a $80$~nm-thick layer of silver is thermally evaporated on top of the structure (deposition parameters: current = $280$~A, deposition-rate = $3$~\r{A}/s).

\underline{Optical setup}

The sample is in between two microscope objectives, one for the excitation and one for detection in transmission configuration. The excitation objective is a $40X$ objective ($N.A.=0.6$) placed at the glass substrate side. The emission and transmission signals are collected by a $100X$ microscope objective ($N.A.=0.9$). The excitation laser for emission measurements is a $3$~eV contiunous-wave laser.
For transmission measurements, polariton modes are resonantly excited by a Xenon white-light source filtered with a $2.4$~eV low-energy pass filter in order to avoid the perovskite crystals degradation.
The propagation measurements (Fig.~2,3 of the main text) are performed with an energy-tunable $50~fs$ pulsed laser ($10~kHz$ repetition rate). The energy selection of the pulsed laser spectrum is made by a double passage through a grating ($1200$~gr/mm) and a slit. After the energy-selection stage, the laser bandwidth is about $2$~meV.


The $100X$ magnified image of the sample surface is projected on a CCD camera.
A second detection path---accessible through a flip mirror (FM)--is used for the detection of the energy dispersion maps. The image on the back focal plane of the detection objective is projected on a spectrometer entrance slit after $0.9X$ magnification. The spectrometer is coupled to a CCD camera allowing for the detection of the polariton energy dispersion. The $50$~cm lens along this path is on a translation stage in order to scan the whole $2D$-momentum space. 
The quarter-wave plate (QWP2) and the half-wave plate (HWP) along the two detection paths allow for the detection of the polarization-resolved maps both in real and momentum space.\\
 \\
 
\underline{Pseudospin extraction} 

The components of the pseudospin, which in our case is the Stokes vector of light, are directly determined by the polarization degrees, measured in 3 possible bases: horizontan/vertical (HV, $s_1$), diagonal/antidiagonal (DA $s_2$), left/right circular (LR, $s_3$). The explicit expressions allowing to obtain the pseudospin from the measured intensities read:
\begin{eqnarray}
    s_1&=&\frac{I_V-I_H}{I_H+I_V}\\
    s_2&=&\frac{I_D-I_A}{I_D+I_A}\nonumber\\
    s_3&=&\frac{I_R-I_L}{I_R+I_L}\nonumber
\end{eqnarray}

\subsection{Supplementary text}

\underline{Rashba spin-orbit coupling as a Yang-Mills field}

In this section, we show explicitly that the Hamiltonian of a massive particle with Rashba spin-orbit coupling is a particular case of a general Hamiltonian of a massive quantum field coupled to the Yang-Mills field. The general relativistic Yang-Mills Lagrangian for the matter field $\phi$ coupled with a non-Abelian field $\bm{F}$ reads (\textit{1},\textit{2}):
\begin{equation}
    \mathcal{L}=\frac{1}{2}\left(D_\mu\bm{\phi}\right)\cdot\left(D^\mu\bm{\phi}\right)-\frac{m^2}{2}\bm{\phi}\cdot\bm{\phi}-\frac{1}{4}\bm{F}_{\mu\nu}\cdot\bm{F}^{\mu\nu}
    \label{LagYM}
\end{equation}
where the covariant derivative contains the coupling with the vector potential $D_\mu\bm{\phi}=\partial_\mu\bm{\phi}+\eta\bm{A}_\mu\times\bm{\phi}$, where the coupling constant  $\eta=\hbar/2$ (the quantum of spin), $\mu,\nu$ span $(t,x,y,z)$ and we use number indices $0-3$ for Pauli matrices. The two first terms are common to any minimally-coupled Lagrangian (note that $D_\mu$ is a covariant derivative, containing the minimally-coupled vector potential $A$). For example, the same terms can be written for the Lagrangian of an electric charge in the electromagnetic field. The form of the last term is specific to the Yang-Mills field (the elements of the field strength tensor $\bm{F}_{\mu\nu}$ are vectors, while for the electromagnetic field they are scalars), it allows to keep the Lagrangian gauge-invariant. The general procedure to obtain the equations of motion for $\phi$ is to use the Euler-Lagrange equations for the matter field $\phi$:
\begin{equation}
    \frac{\partial L}{\partial\phi}-\frac{\partial}{\partial x^\mu}\frac{\partial L}{\partial (D_\mu\phi)}=0
    \label{EuLag}
\end{equation}
Since we are dealing with non-relativistic particles, we can first write a non-relativistic version of the Lagrangian~(S1), separating the spatial and temporal parts of the four-vectors:
\begin{eqnarray}
    L_{NR}&=&\frac{i\hbar}{2}\left(\dot\phi^\dagger\phi -\phi^\dagger\dot\phi\right)+\phi^\dagger\eta A_0^a\phi\\
    &+&\frac{1}{2m}\left[\left(\hat{\bm{p}}-\eta\bm{A}^a\sigma^a\right)\psi\right]^\dagger\left[\left(\hat{\bm{p}}-\eta\bm{A}^a\sigma^a\right)\psi\right]\nonumber\\
    &-&\frac{1}{4}F_{\mu\nu}^a F_{\mu\nu}^a
\end{eqnarray}
from which, using the Euler-Lagrange equations~(S2), one obtains the following Schrodinger equation for a spinor $\phi$:
\begin{equation}
    i\hbar\frac{\partial \phi}{\partial t}=\left[{ \frac{1}{2m}\left(\hat{\bm{p}}-\eta\bm{A}^a\sigma^a\right)^2}+\eta A_0^a\sigma^a\right]\phi
\end{equation}
This is the most general form of the Schrodinger equation for the wavefunction of a particle with a vectorial charge (for example, spin), coupled to an arbitrary field acting on this charge. The expression in the parenthesis in the right-hand part of this equation can be directly compared with the Rashba Hamiltonian from the main text
\begin{equation}
    \hat{H}_R=\frac{1}{2m}\left(\hat{\mathbf{p}}+m\alpha \mathbf{\sigma}\right)^2
    \label{hamR2}
\end{equation}
allowing to identify the components of the vector potential $A_x^1=A_y^2=-m\alpha/\eta$,  as stated in the main text. We can therefore conclude that the Rashba spin-orbit coupling represents a particular case of the Yang-Mills field with two non-zero constant components of the vector potential.

The remaining term of the equation (S5), $A_0^a\sigma^a$ corresponds to the Zeeman splitting, which can appear in the presence of an applied magnetic field (\textit{22}). This field affects the phase of the wave function and it can rotate the spin, but does not couple with the spatial degrees of motion.

\underline{The Rashba spin-orbit coupling and the Dirac equation}

In several recent works (\textit{14},\textit{15}), the Rashba-like spin-orbit coupling is used to construct a 2D massive Dirac Hamiltonian. While this Hamiltonian indeed contains non-commuting (and thus non-Abelian) Pauli matrices, it should not be considered as an example of a non-Abelian \emph{gauge field}, because the 2D Dirac equation by itself does not describe a charge coupled to a field, but rather a freely propagating particle. Indeed, the basic 2D Dirac Hamiltonian with a mass term reads
\begin{equation}
    \hat{H}_{D}=\hbar c \left(\sigma_x k_x+\sigma_y k_y\right)+\sigma_z mc^2= c \bm{\sigma}\cdot\bm{p}+\sigma_z mc^2
\end{equation}
The corresponding Hamiltonian is not in a minimal coupling form describing the interaction between a charge and a vector potential $\bm{A}$, because it describes a free particle.
Then, the 2D Dirac Hamiltonian for a charged particle in presence of a vector potential reads:
\begin{equation}
    i\hbar\frac{\partial\psi}{\partial t}= c \bm{\sigma}\cdot\left(\bm{p}-\frac{e}{c}\bm{A}\right)+\sigma_z mc^2
\end{equation}
This is how the gauge field (the electromagnetic field in this case) should be introduced in the Dirac equation. In this particular case, the gauge field is Abelian. A non-Abelian field can be introduced in a similar way.

\begin{figure}[tbp]
\centering{}\includegraphics[scale=0.4]{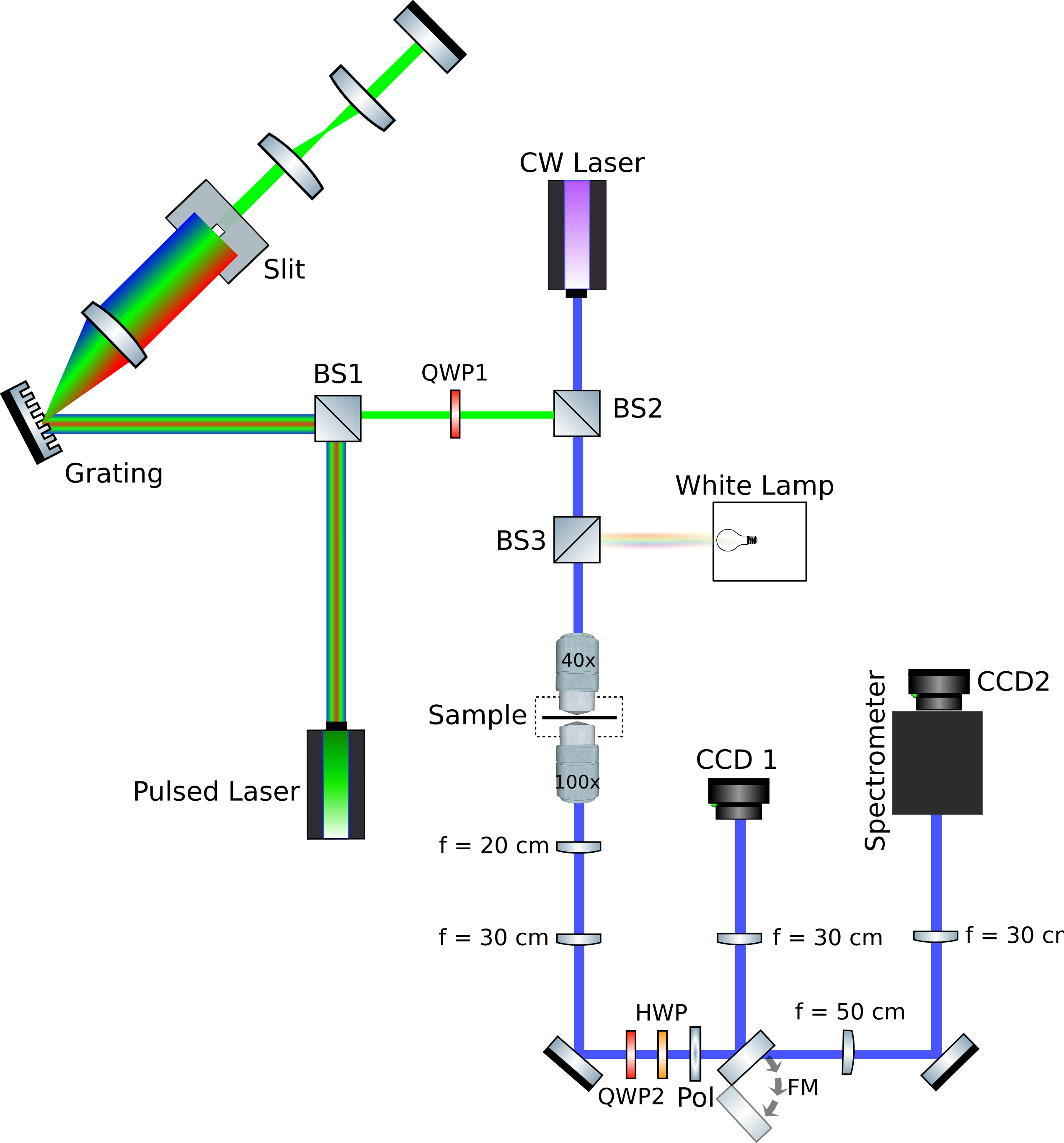}
\caption{Sketch of the optical setup.}
\end{figure}

\end{document}